\documentclass[amssymb,prb,twocolumn,showpacs]{revtex4}
\usepackage{epsfig}
\usepackage{dcolumn}
\usepackage{amsmath}
\usepackage{epsfig}
\let\up=\uparrow%
\let\down=\downarrow%
\begin{document}

\title{Andreev drag effect in ferromagnetic-normal-superconducting systems} 
\author{David S\'anchez}
\affiliation{D\'epartement de Physique Th\'eorique,
Universit\'e de Gen\`eve, CH-1211 Gen\`eve 4, Switzerland}
\author{Rosa L\'opez}
\affiliation{D\'epartement de Physique Th\'eorique,
Universit\'e de Gen\`eve, CH-1211 Gen\`eve 4, Switzerland}
\author{Peter Samuelsson}
\affiliation{D\'epartement de Physique Th\'eorique,
Universit\'e de Gen\`eve, CH-1211 Gen\`eve 4, Switzerland}
\author{Markus B\"uttiker}
\affiliation{D\'epartement de Physique Th\'eorique,
Universit\'e de Gen\`eve, CH-1211 Gen\`eve 4, Switzerland}
\date{\today}

\begin{abstract}
We investigate conductances and current correlations in a system 
consisting of a normal multichannel conductor connected to one superconducting
and two ferromagnetic electrodes. For antiparallel orientation of the 
ferromagnet polarizations, current injection from one ferromagnet
can, due to Andreev reflection,
lead to a net drag of current from the second ferromagnet
toward the superconductor. We present the 
conditions for the Andreev drag in terms of lead
polarizations, contact conductances and spin-flip scattering.
Remarkably, both equilibrium and
nonequilibrium zero-frequency
current correlations between the ferromagnets
become positive even in the presence of spin relaxation.
\end{abstract}


\maketitle

\section{Introduction}
At a normal-superconducting
interface, Andreev reflection causes a conversion of
the quasiparticle charge \emph{and} a flipping of its spin---an
electron with spin up, incoming towards the  superconductor,
is reflected as a hole with spin down.
This leads to exciting consequences in the transport physics
where the focus is put on creating, manipulating and detecting
spin currents and magnetization.\cite{wol01,fie99}
When the carriers are spin polarized, the transport properties
are strongly sensitive to the relative orientation
of the ordered moments of ferromagnetic electrodes,\cite{bai88,slo89}
a feature that has received a lot of interest
due to the relatively long spin dephasing times observed
in metals\cite{joh85,jed01} and  semiconductors.\cite{kik97} 
In hybrid systems containing a superconductor, various aspects
of the interplay between Andreev reflection
and ferromagnetism have been addressed
theoretically\cite{theoryFM,bel00}
and experimentally.\cite{expFM,expFM2,expFM3}

\begin{figure}[b]
\centerline{
\epsfig{file=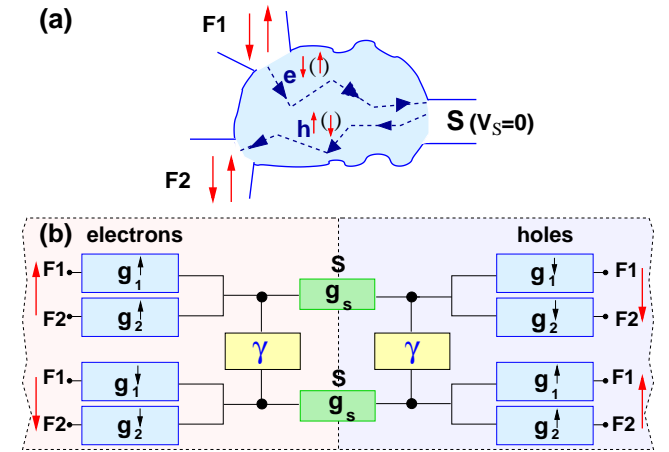,angle=0,width=0.47\textwidth,clip}
}
\caption{(a) Generic ferromagnetic-normal-superconducting
structure: the process that gives rise to the Andreev
drag effect is indicated.
(b) Sketch of the equivalent circuit model (see text).
}
\label{fig1}
\end{figure}

The structure of interest here is shown in Fig.~\ref{fig1}(a):
two ferromagnetic contacts are connected to a normal conductor
which is in turn connected to a superconductor.
Of particular interest is the case of crossed Andreev reflection 
between ferromagnetic leads with antiparallel
polarization dominating over crossed normal reflection.
In such a geometry, with a voltage applied to one 
of the leads while the other lead and the superconductor are grounded,
the injected current from the biased ferromagnetic lead 
effectively {\em drags} along a current from the grounded lead. 
This effect is easy to understand since the transfer of Cooper pairs 
into the superconductor requires a pair of electrons with opposite spin. 
If the injected current consists only of, say, spin-up carriers,
the system supplies the spin-down carriers from the
oppositely polarized ferromagnetic lead even though 
no voltage is applied at this contact.

The physics of the Andreev drag effect
was already addressed by 
Jedema and van Wees {\em et al.}\cite{jed99} for a multi-terminal 
metallic diffusive conductor. These authors pointed to negative 
four-terminal resistances as a consequence of the Andreev drag effect. 
In contrast, much of the recent theoretical discussion has focused 
on a geometry where two 
needle-shaped ferromagnetic leads with antiparallel 
polarization are directly coupled to a superconductor 
at two spatially separated points.\cite{deu00}
In this case, even in the 
case of $100\%$ spin polarization, the crossed Andreev 
reflection probability is strongly suppressed
in realistic geometries,\cite{cho00}
making the experimental detection of the drag effect 
difficult. In addition, the direct connection of the
ferromagnetic electrodes to the superconductor
might lead to unwanted modifications of 
the local properties of the ferromagnet\cite{expFM2} and
the superconductor,\cite{expFM3} suppressing the drag effect. 

In this work, we overcome the above difficulties by inserting 
a normal multichannel\cite{les01} mesoscopic conductor
with generic elastic scattering between 
the ferromagnetic electrodes and the superconductor [see the sketch of 
Fig.~\ref{fig1}(a)].  We include
arbitrary interface conductances and spin-flip relaxation (see below).
The advantage of this scheme is twofold:
(i) since the superconductor is only contacted at one 
point, the drag effect develops to the fullest extent possible,
being of the order of the conductances
of the contacts between the normal conductor 
and the ferromagnetic and superconducting electrodes;
and (ii) the sources of the competing phenomena---superconductivity and 
ferromagnetism---are spatially separated by the normal conductor.

We find that the Andreev drag effect is observable
when the conductance of the normal-superconducting (NS) interface
dominates over the ferromagnetic-normal (FN) contact conductances.
The drag effect is reinforced with increasing lead polarization
and survives for a considerable amount of spin-flip scattering.
Furthermore, we demonstrate that the crossed Andreev reflections
have a profound influence on the
zero-frequency current--current cross-correlations.
We observe that both the thermal noise and the shot noise power
measured between the two ferromagnetic leads
yield \emph{positive} values. 

\section{Model}
\label{sec-mod}
We analyze a hybrid system (shown in Fig.~\ref{fig1}) 
consisting of a normal conductor connected to two ferromagnetic 
(F1 and F2) and one superconducting (S) reservoirs.
For simplicity, the direction of 
the magnetization of the two ferromagnets is assumed to be collinear
and we consider identical FN couplings.
Any type of contacts,
e.g. diffusive-, ballistic- or tunnel contacts, can be 
treated. In what follows, we deal
with point contacts at the FN
(NS) interface with $N$ ($M$) transversal modes and mode independent 
transparency $\Gamma$ ($\Gamma_S$). We consider the case with dimensionless
conductances (in units of $2e^2/h$) much larger than unity:
$g=N\Gamma,g_S=M R \gg 1$, where $R=\Gamma_S^2/(2-\Gamma_S)^2$
is the Andreev reflection probability.
Under these conditions, weak localization as well as 
Coulomb blockade effects can be safely neglected.
Moreover, the dwell time of the normal conductor, $\tau_d$, is
assumed to be shorter than typical inelastic scattering times.

Throughout the paper, we consider the case when the proximity
effect inside the normal conductor is suppressed
(but the Andreev reflection at the NS interface
is taken into account).
A strong proximity effect would quench the drag effect
as the normal conductor would behave effectively as a superconductor.
\cite{note}
This can be avoided either with a magnetic field
(e.g., a stray field from the ferromagnets
or an externally applied field) or with a characteristic
quasiparticle energy (bias $eV$ or temperature $k_B T$)
much larger than the inverse dwell time,
$e V, k_B T \gg \hbar/\tau_{d}$.
Nevertheless, to avoid single particle transport,
both $e V$ and $k_B T$ should be smaller than the gap
of the superconductor. 

The resistance of the system is completely dominated by the 
resistances of the FN and NS contacts.
Moreover, we assume that the physical 
properties within the conductor
are isotropic due to diffusive or chaotic scattering. 
We can thus consider the normal 
conductor to be effectively zero dimensional. This restricts our model 
to systems with a spin-flip length much larger than the lateral 
dimensions of the system, i.e., there are no spatial variations of 
accumulated spins inside the dot, though no 
assumptions are made about the relation between the spin-flip time 
$\tau_{\rm sf}$ and $\tau_d$.
We stress that all the above constraints form a situation
that is experimentally accessible with present techniques.
Vertical structures like metallic thin-film heterostructures,
epitaxial layers of half-metallic chromium dioxide
or diluted magnetic semiconductors
are equally suitable for the observation of the effects
discussed here. In fact, a uniform spin accumulation
was very recently achieved in a normal metallic island
coupled to magnetic leads.\cite{zaf03}

\begin{figure}[t]
\centerline{
\epsfig{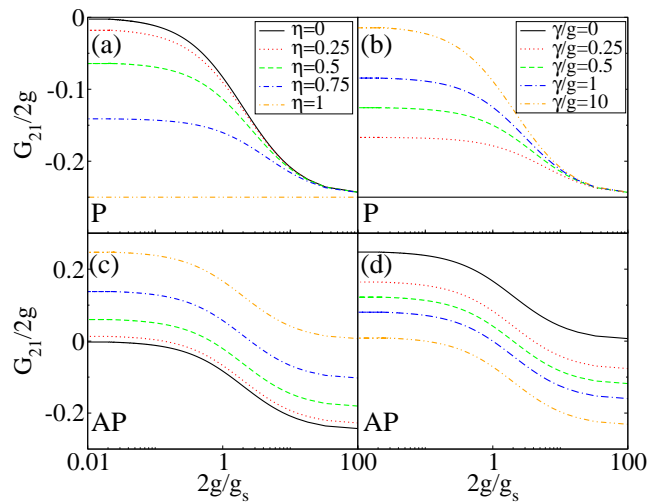}
}
\caption{Off-diagonal conductance $G_{2 1}=(h/2e^2) \mathcal{I}_2/V_1$
of the current in ferromagnetic lead F2 when
a voltage bias is applied to lead F1 as a function of
the contact conductances. The magnetic moments of the reservoirs
are aligned parallel in (a) with spin-flip rate $\gamma=0$ and
(b) with polarization $\eta=1$. (c) and (d) same as
(a) and (b) in the antiparallel orientation.
}
\label{fig2}
\end{figure}

General theoretical formulations of semiclassical spin transport,
based on quasiclassical Green's functions, have been developed
by Brataas, Nazarov and Bauer for normal-ferromagnetic
systems\cite{bra00} and
by Belzig {\em et al.} for normal-ferromagnetic-superconducting
systems.\cite{bel00} Here, we follow the approach
of Samuelsson and B\"uttiker~\cite{sam02}
extended to spin-dependent transport.
This simple theory is equivalent to Refs.~\onlinecite{bel00,bra00}
for the geometry in Fig~\ref{fig1}(a). It is directly
formulated in terms of distribution functions and allows us
to treat the current and the current correlations
within a unified framework.
Because the proximity effect is suppressed, the electrons and holes
are independent~\cite{sam02} and the system is mapped onto
the equivalent circuit of Fig.~\ref{fig1}(b) in which
the electron and hole subsystems
(with opposite spins) are connected via
Andreev reflection. Spin relaxation in the normal 
conductor gives rise to a spin-flip current between the spin 
subsystems with the same quasiparticle charge.
As a result, we can describe the
conductor with four nodes [two nodes ($\up$ and $\down$)
for each quasiparticle type (electrons and holes)]
connected to each other and to the
ferromagnetic reservoirs via spin dependent resistances.
We model the spin dependent conductances of the FN contacts
as $g_{\alpha}^{\up,\down}=g (1\pm\eta_{\alpha})/2$,
where $\eta_\alpha$ is the spin polarization of the lead $\alpha$
($\alpha=\{1,2\}$): $\eta_\alpha=(\nu_{\alpha}^{\up}-\nu_{\alpha}^{\down})/ 
(\nu_{\alpha}^{\up}+\nu_{\alpha}^{\down})$,
where $\nu_{\alpha}^{\sigma}$ is the density of states of
the ferromagnetic lead close to the Fermi energy ($\sigma=\{\up,\down\}$).

Conservation of time-averaged (spin-dependent)
current densities must 
be fulfilled at each node for every energy $E$.
Currents flowing into the normal conductor are defined positive
such that the {\it spectral} currents obey 
\begin{equation} \label{eq-circuit} 
\sum_\alpha i_{\alpha q}^\up \mp i_{S}^\up 
+i_{\down\up,e}=0\,, \,\,  
\sum_\alpha i_{\alpha q}^\down \mp i_{S}^\down 
-i_{\down\up,e}=0 \,, 
\end{equation} 
where $q=(e,h)$ labels the carrier (electron or hole)
and $-$ ($+$) is for $q=e$ ($q=h$).
The currents through the FN and NS contacts are: 
\begin{equation} 
i_{\alpha q}^\sigma=g_\alpha^\sigma 
(f_{\alpha q}-\hat{f}_{q}^\sigma), \,\,\,\, 
i_{S}^\sigma=g_S (\hat{f}_{e}^{\bar{\sigma}}-\hat{f}_{h}^\sigma)/2 \,, 
\label{eq-curr} 
\end{equation} 
where $f_{\alpha q}(E)=\{ 1+\exp{([E\mp eV_\alpha]/kT)} \}^{-1}$ is 
the equilibrium distribution function at reservoir $\alpha$. Note that 
$f_{\alpha q}$ is spin independent as spin relaxation mechanisms are 
usually highly efficient in ferromagnets.
In Eq.~(\ref{eq-curr}), $\hat{f}_{q}^\sigma (E)$
is the nonequilibrium distribution function of the normal 
conductor to be calculated from Eq.~(\ref{eq-circuit}).
The spin-flip current
$i_{\down\up,q}=\gamma (\hat{f}_{q}^\down -\hat{f}_{q}^\up)$
equilibrates the spin distributions. Here, 
$\gamma=h\nu_0/\tau_{\rm sf}$ ($\nu_0$ is the density of states of 
the conductor at the Fermi energy
and $\tau_{\rm sf}$ the spin-flip time).\cite{bra00}

\section{Conductance matrix}
\label{sec-cm}
The {\it total} current
flowing out of reservoir $\alpha$ is
\begin{equation}
\mathcal{I}_{\alpha}=\mathcal{I}_{\alpha e}^\uparrow+
\mathcal{I}_{\alpha e}^\downarrow- \mathcal{I}_{\alpha h}^\uparrow-
\mathcal{I}_{\alpha h}^\downarrow\,,
\end{equation}
where $\mathcal{I}_{\alpha q}^\sigma=(e/h)
\int dE \, i_{\alpha q}^\sigma(E)$,
from which we determine the conductance matrix.
The conductance $G_{\alpha \beta}$ relates the current
$\mathcal{I}_{\alpha}$ with a voltage change at lead $\beta$.
In the case $V_1>0$ and $V_2=V_S=0$, the interesting
conductance is $G_{2 1}=(h/2e^2)(\mathcal{I}_2/ V_1)$:
\begin{equation}
G_{2 1}=\frac{g^2 [ g ( {\eta_1}^2 + {\eta_2}^2 -2 )
- 4 \gamma - 2 \eta_1 \eta_2 g_S ] }{g^2 
[ 4 - (\eta_1+ \eta_2)^2 ]
+ 8 \gamma g_S + 4 g \left( 2 \gamma + g_S \right) }\,.
\end{equation}
In a normal system, $G_{2 1}$ is always negative.\cite{but86}
On the contrary, in our system we can have $G_{2 1}>0$,
which is the manifestation of the Andreev drag effect.\cite{pha02}
Figure~\ref{fig2} shows the behavior of $G_{2 1}$ as a function
of $2g/g_S$. In the parallel (P) case, $\eta_1=\eta_2=\eta$
and $G_{2 1}$ is negative [Fig.~\ref{fig2}(a)].
The maximum (absolute) value of $G_{2 1}$ is attained at $\eta=1$
and decreases with increasing values of $\gamma$ [Fig.~\ref{fig2}(b)].
In the antiparallel (AP) configuration ($\eta_1=-\eta_2=\eta$),
a \emph{reversal} of the current at F2 occurs
for low values of $g/g_S$ and $\gamma$
[see Fig.~\ref{fig2}(c) and Fig.~\ref{fig2}(d)].
Note that the drag effect can result already
for a small value of $\eta$.
In fact, crossed Andreev reflections start to dominate
when $\eta^2>(g/g_S)/ (1+g/g_S)$ at $\gamma=0$.
In particular, for the fully polarized case
one always finds $G_{2 1}>0$
\emph{regardless of the ratio $g/g_S$}.
When $g_S\gg g$, we find that $G_{2 1}>0$ holds for any value of $\eta>0$,
in agreement with Ref.~\onlinecite{jed99}.
Increasing $\gamma$, the spin relaxation processes
randomize the spin distributions and eventually lead to $G_{2 1}<0$,
[the curves tend to the $\eta=0$ case of Fig.~\ref{fig2}(b)].
The condition for the drag effect is:
\begin{equation}
2\gamma < \eta^2 g_S-g(1-\eta^2)\,.
\end{equation}
This is a central result of our work.

Interestingly, the Andreev drag effect
implies that even the equilibrium Nyquist-Johnson
noise yields \emph{positive} current cross-correlations
$P_{2 1}=4 k_B T G_{2 1}$ in contrast to normal conductors.
Nevertheless, energy dissipation is ensured
as the eigenvalues of the conductance matrix are semipositive.

Another striking result is that for $\eta=1$ there are
two linearly independent voltage configurations that yield
\emph{zero} eigenvalue of the conductance matrix.
One is the gauge solution ($V_1=V_2=V_S$).
The nontrivial arrangement is $V_1=-V_2=V$ and $V_S=0$.
Here, F1 injects electrons whereas an equivalent flow
of incident holes stems from F2.
Complete Andreev reflection yields then zero net current.

\section{Magnetization}
\label{sec-mag}
Out of equilibrium for $V_1>0$ and $V_2=V_S=0$,
an accumulation of majority spins forms within the conductor.
Before applying the voltage, the conductor
has zero magnetization. In the AP
case, the polarization excess is given by
\begin{equation}
M=\frac{\int dE \sum_{q}(\hat{f}_{q}^\up-\hat{f}_{q}^{\down})}
{\int dE \sum_{q}(\hat{f}_{q}^\up+\hat{f}_{q}^{\down})}
=\frac{g\eta}{2\gamma+g+g_S}\,.
\end{equation}
It approaches $\eta$ for $g\gg g_S,\gamma$
and vanishes for $\gamma\gg g$ or $g_S\gg g$. 

\begin{figure}[t]
\centerline{
\epsfig{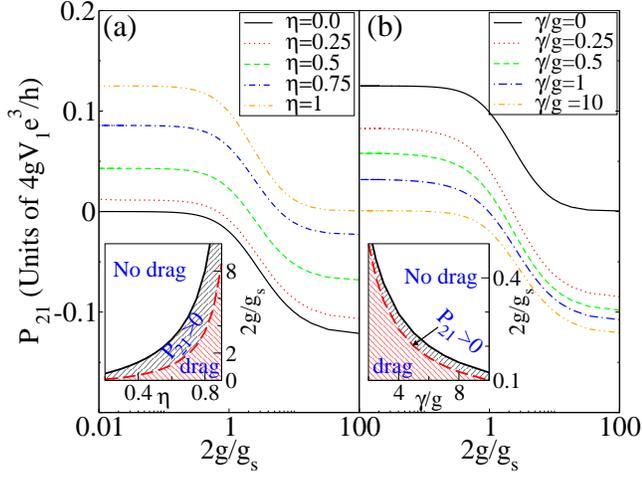}
}
\caption{Cross correlation $P_{1 2}$ for $\Gamma=1$ and $R=1$
versus the ratio of the conductances of the point contacts
in the antiparallel case.
In (a) $\gamma=0$ whereas in (b) $\eta=1$.
Insets: ``Phase diagram'' of the occurrence of the Andreev
drag effect (below the dashed line)
and positive $P_{1 2}$ (below the solid line)
for (a) $\gamma=0$ and (b) $\eta=1$.
}
\label{fig3}
\end{figure}

\section{Cross correlations}
\label{sec-cc}
To obtain additional insight into the 
drag effect, we investigate the shot noise of the current.
Of particular interest are the correlations between currents
flowing in the two ferromagnetic reservoirs.
In purely normal systems, the current cross-correlations
are manifestly negative.\cite{but92} However, in 
hybrid superconducting structures it was shown that Andreev 
reflections can cause a reversal of the sign of the cross-correlations.
\cite{Poscorr} The positive correlations in few-mode conductors
were also shown to be 
enhanced by ferromagnetic leads.\cite{tad02} Here, we focus on the
cross correlations in the AP configuration, which is the 
most interesting case.

The contacts act as generators of fluctuations 
$\delta i_{\alpha q}^\sigma$ and $\delta i_{S}^\sigma$.
As a result, the spin-dependent 
distribution function of the normal conductor fluctuates itself. The 
total fluctuating current is
\begin{equation}
\Delta i_{\alpha q}^\sigma = \delta i_{\alpha q}^\sigma-
g_{\alpha}^\sigma \delta \hat{f}^{\sigma}_{q}, \,\,\,\,
\Delta i_{S}^\sigma = \delta i_{S}^\sigma+
g_{S} (\delta \hat{f}_{e}^{\bar{\sigma}}-\delta \hat{f}_{h}^{\sigma}) \,.
\label{eq-fluct}
\end{equation}
We also take into account
the fluctuations $\delta i_{\down\up,q}$
due to spin relaxation, giving the fluctuating current
\begin{equation}
\Delta i_{\down\up,q} = \delta i_{\down\up,q}
+\gamma (\delta \hat{f}^{\down}_{q}-\delta \hat{f}^{\up}_{q})\,.
\end{equation}

The next step is to determine the fluctuations of the
distribution functions of the conductor,
$\delta \hat{f}^{\sigma}_{q}$. In the zero frequency limit, the 
fluctuating currents are conserved ($\mp$ for $q=e,h$):
\begin{subequations}
\label{eq-fluctcircuit}
\begin{eqnarray}
\sum_\alpha\Delta i_{\alpha q}^\up\mp\Delta i_{S}^\up
+\Delta i_{\down\up,q}=0\,, &\\
\sum_\alpha\Delta i_{\alpha q}^\down\mp\Delta i_{S}^\down
-\Delta i_{\down\up,q}=0\,. &
\end{eqnarray}
\end{subequations}
The fluctuations of the charge current
at contact $\alpha$ are found from $\Delta i_{\alpha}=
\Delta i_{\alpha e}^\up+\Delta i_{\alpha e}^\down-
\Delta i_{\alpha h}^\up-\Delta i_{\alpha h}^\down $.

The noise power of the current cross-correlations is
\begin{equation}
P_{2 1}=2\int dt \,
\langle\Delta{\mathcal{I}_2}(t)\Delta{\mathcal{I}_1}(0)\rangle \,.
\end{equation}
The contacts emit fluctuations independently so that
\begin{equation}
\langle \delta i_{\alpha q}^\sigma (E,t )
\delta i_{\alpha' q'}^{\sigma '} (E',t') \rangle =
\frac{h}{e} \bar{\delta}
S_{\alpha q}^\sigma\,,
\end{equation}
with $\bar{\delta}\equiv \delta_{\alpha \alpha'}\delta_{q q'}
\delta_{\sigma\sigma '} \delta (E-E') \delta (t-t')$. Here,
$S_{\alpha q}^\sigma$ is the fluctuation power:\cite{sam02,but92} 
\begin{subequations}
\begin{eqnarray}
&S_{\alpha q}^\sigma=e g_{\alpha}^{\sigma}
[f_{\alpha q}^\sigma+\hat{f}_{q}^\sigma -2 \hat{f}_{q}^\sigma f_{\alpha q}^\sigma-
\Gamma (f_{\alpha q}^\sigma -\hat f_{q}^\sigma )^2 ]\,,  \\
&S_{S}^\sigma=e g_S [\hat{f}_{e}^\sigma+
\hat{f}_{h}^{\bar{\sigma}}- 2\hat{f}_{h}^{\bar{\sigma}}\hat{f}_{e}^\sigma-
R (\hat{f}_{e}^\sigma -\hat{f}_{h}^{\bar{\sigma}} )^2 ]/2\,, &\\
&S_{\down\up q}=e\gamma [\hat{f}_{q}^\down
+\hat{f}_{q}^\up-2\hat{f}_{q}^\up\hat{f}_{q}^\down ] \,,
\end{eqnarray}
\end{subequations}
where we have assumed that the statistics
of spin-flip events is Poissonian.\cite{mis03}

The above system of equations is complete and can be solved
for any set of applied voltages, polarization alignments
and spin-flip rates. We calculate $P_{2 1}$ for the bias configuration
given by $V_1>0$, $V_2=V_S=0$ in the AP case at $k_B T=0$. 
We present only the analytical result
in limiting cases and show the general result in Fig.~\ref{fig3}.
In Fig.~\ref{fig3}(a), we plot $P_{2 1}$
as a function of the conductance ratio $2g/g_S$
for various polarizations $\eta$, $\gamma=0$ and $\Gamma=R=1$.
(Qualitatively, the results are independent of $\Gamma$ and $R$.)
For $\eta=0$, the cross-correlations
are negative. As $\eta$ is turned on, $P_{2 1}$ can become positive.
With increasing $g/g_S$, $P_{2 1}$ exhibits a crossover
from negative to positive
values at $2g/g_S\sim 1$.
Indeed, for dominating NS conductance (small $g/g_S$)
we find
\begin{equation}
P_{2 1}=p_0 \eta^2
[\Gamma+2-\eta^2(3\Gamma-2)]/16+\mathcal{O}(g/g_S) \,,
\end{equation}
which is manifestly positive for $\eta>0$
($p_0\equiv 4gV_1 e^3/h$).
In this case, the electron and hole distributions
within the conductor are almost identical.

On the other hand, for dominating FN conductance (small $g_S/g$)
the cross-correlations become
\begin{equation}
P_{2 1}=p_0 (1-\eta^2) [\Gamma-2+\eta^2(3\Gamma-2)]/8
+\mathcal {O}(g_S/g)\,,
\end{equation}
which are negative for $\eta<1$.
Notice that for $\eta=0$ this expression yields
the shot noise for a chaotic dot coupled
to two normal leads.
In the fully polarized case, the cross-correlations
are \emph{always} positive independently of $g/g_S$.

For illustrative purposes we show in the inset of Fig.~\ref{fig3}(a)
the ``phase diagram'' $(\eta,g/g_S)$. The solid line marks
the crossover between positive and negative cross-correlations.
With increasing $\eta$, the crossover point shifts toward
larger values of $2g/g_S$.
For comparison, we also plot a dashed curve that marks the
transition to the Andreev drag effect.
Notice that the conditions $P_{2 1}>0$ and $G_{2 1}>0$ 
are closely related, but not identical. This is the case 
since the scattering processes in general contribute in
different ways for the current and noise.\cite{sam02}

A finite source of spin relaxation leads to the suppression
of positive $P_{2 1}$. To see this, we show in Fig.~\ref{fig3}(b) the
cross-correlations for the fully polarized case for
different values of $\gamma$. For nonzero values of $\gamma$,
$P_{2 1}$ becomes negative with increasing $2g/g_S$.
Eventually, when $\gamma\to\infty$, we recover the unpolarized
case ($\eta=0$) of Fig.~\ref{fig3}(a).

\section{Conclusions}
\label{sec-con}
We have investigated the transport properties
of a normal conductor attached to ferromagnetic leads and coupled
to a superconductor in the subgap regime.
The proposed geometry, which includes
generic elastic and spin-flip scattering,
is shown to exhibit a pronounced Andreev drag effect.
We have demonstrated that
the equilibrium and transport
current correlations are strongly sensitive 
to crossed Andreev reflections. 
The theory presented here applies to a wide range of geometries
and should therefore advance experimental observation of the 
Andreev drag effect. 

{\it Note added in proof}. Recently, we became aware of the work
of Lambert {\it et al.}\cite{lam03} who compute numerically
the conductances for a metallic diffusive geometry similar
to Fig.~1. They also conclude that a normal conductor inserted
between the ferromagnetic and the superconducting contacts
leads to an enhancement of the Andreev drag effect. In contrast
to our work, the role of spin-flip, the effect of differing
lead polarizations, and the shot noise are not investigated.

\section*{Acknowledgments}
We thank V.~Chandrasekhar and B.J.~van~Wees for helpful discussions.
We acknowledge support from the Swiss NSF through the program MANEP.
D.S. and R.L. were also supported by the EU
RTN under Contract No. HPRN-CT-2000-00144, Nanoscale Dynamics.


\begin{thebibliography}{90}
\bibitem{note}
We are then back to the situation of Ref.~\onlinecite{deu00}.
\end{thebibliography}
\end{document}